\documentclass[trackchanges,twocolumn]{aastex701}

\shorttitle{Optical Follow-Up of Two Spider Pulsar Candidates}
\shortauthors{B. Due, J. Bridziute, M. Linares, M. Turchetta, M. Satybaldiev}

\begin{document}
\title{Optical Follow-Up of Two Spider Pulsar Candidates:\\ An Irradiated Redback and a Possible Double Giant Impostor}

\author[orcid=0009-0001-0330-6086, gname=Bettina, sname=Due]{Bettina Due}
\affiliation{Department of Physics, Norwegian University of Science and Technology, NO-7491 Trondheim, Norway}
\email{bettina.due8@gmail.com}  

\author[gname=Jundile, sname=Bridziute]{Jundile Bridziute}
\affiliation{Department of Physics, Norwegian University of Science and Technology, NO-7491 Trondheim, Norway}
\email{jundileb@stud.ntnu.no} 

\author[orcid=0000-0002-0237-1636, gname=Manuel, sname=Linares]{Manuel Linares} 
\affiliation{Department of Physics, Norwegian University of Science and Technology, NO-7491 Trondheim, Norway}
\affiliation{Departament de Física, EEBE, Universitat Politècnica de Catalunya, Av. Eduard Maristany 16, E-08019 Barcelona, Spain}
\email{manuel.linares@ntnu.no}

\author[orcid=0000-0003-0438-4956, gname=Marco, sname=Turchetta]{Marco Turchetta}
\affiliation{Department of Physics, Norwegian University of Science and Technology, NO-7491 Trondheim, Norway}
\email{marco.turchetta@ntnu.no}

\author[orcid=0009-0001-3795-4049, gname=Maksat, sname=Satybaldiev]{Maksat Satybaldiev}
\affiliation{Department of Physics, Norwegian University of Science and Technology, NO-7491 Trondheim, Norway}
\email{maksat.satybaldiev@ntnu.no}

\begin{abstract}

Compact binary millisecond pulsars (nicknamed ``spiders") are a growing class of neutron stars: more than 100 systems are known in the Galactic field, including more than 30 candidates where pulsations have not been detected. Here we report the results of an optical follow-up of two spider pulsar candidates: 2FGL~J0846.0+2820 (J0846, a huntsman candidate in an 8.1~d orbit) and 4FGL~J0935.3+0901 (J0935, with a 2.4~h orbital period and proposed as a spider of either the transitional, black widow or redback sub-class). We also analyzed their multi-wavelength properties (infrared, X-ray and $\gamma$-ray bands), in order to assess their nature and current state. The optical orbital light curves of J0846 change from double-peaked in 2023 to single-peaked in 2024, but we argue they are not consistent with either irradiation-driven or ellipsoidal orbital modulation. Its spectral energy distribution shows evidence for two blackbody-like components with similar temperatures ($T$) and radii ($R$), which dominate the optical ($T_2\simeq6000$~K, $R_2\simeq1.8$~R$_\odot$) and infrared ($T_1\simeq3400$~K, $R_1\simeq5.2$~R$_\odot$) bands. Based on this, we argue that J0846 is not a bona fide pulsar candidate and propose an alternative identification: a detached double red giant. In J0935, we find a light curve with one flux maximum around superior conjunction of the companion, where the colors and inferred temperature increase, suggesting prominent irradiation from the pulsar. We place constraints on its companion base temperature ($4000\!-\!4600$~K), which support one of the three classifications proposed for J0935: a redback spider in the pulsar state.

\end{abstract}


\begin{center}
    {\it Submitted to ApJ}
\end{center}

\section{Introduction} \label{sec:introduction}

Millisecond pulsars (MSPs) are formed when neutron stars (NSs) are spun up to millisecond spin periods within low-mass X-ray binaries (LMXBs; \citealt{Alpar1982}). LMXBs host old, weakly magnetized NSs in orbit with low-mass companions. The NS accretes matter from the companion, powering luminous X-ray emission during the LMXB phase, after which rotation-powered pulsed emission --- mostly in the $\gamma$-ray and radio bands --- is turned on again \citep{Tauris2013}. Compact binary MSPs, also known as ``spiders", have non-degenerate companions in tight orbits (with orbital periods typically $P_\mathrm{orb}\lesssim 1$~d). Among spiders, three systems known as transitional MSPs have been observed to switch between an accretion-powered disc state and a rotation-powered radio pulsar state, providing strong evidence for this so-called recycling scenario \citep{Archibald2009, Papitto2013, Bassa2014}.

Spiders are commonly divided into redbacks (RBs), with low-mass non-degenerate companions of $\sim\!0.1\!-\!0.7~\mathrm{M}_\odot$ \citep{Roberts2013}, and black widows (BWs), with very low-mass semi-degenerate companions of $<\!0.1~\mathrm{M}_\odot$ \citep{Fruchter1988}. Due to their compact orbits, spider companions can be gradually ablated by the relativistic pulsar wind \citep{Heuvel1988}, inspiring their cannibalistic nicknames. The MSP radio signal is absorbed and dispersed by intra-binary material expelled from the companion star, producing long low-frequency eclipses \citep{DAmico2001, Polzin2020}. The interaction between the pulsar and companion winds usually forms an intra-binary shock (\citealt{Phinney1988}). A third category of spiders has recently emerged with the discovery of PSR~J1417--4402 (J1417) \citep{Strader2015, Camilo2016} and PSR~J1947--1120 (J1947) \citep{Strader2025}, known as huntsman spiders. They are characterized by longer $\sim\! 2\!-\!10 ~\mathrm{d} $ orbital periods, and host giant companions in a mass range similar to that of RBs. Their evolutionary track may also differ from RBs and BWs; huntsman secondaries could be in the ``red bump" phase, during which mass transfer is temporarily halted as the giant companions underfill their Roche lobes \citep{Strader2025}.

The long mass-accretion phase preceding their formation makes spiders promising sites to find supermassive NSs exceeding $2~\mathrm{M}_\odot$ \citep{Linares2020}. Among the $>\!500$~MSPs identified in the Galactic field \citep{Bhattacharyya2022}, 86 Galactic spiders and 36 spider candidates have been cataloged to date based on their multi-wavelength properties (\citealt{Koljonen2025}\footnote{SpiderCat v1.9.5; \url{https://astro.phys.ntnu.no/SpiderCAT}}). The known population of spiders has grown substantially since the launch of the Fermi Large Area Telescope ({\it Fermi}-LAT; \citealt{Atwood2009}), allowing for the discovery of radio and $\gamma$-ray pulsars thanks to multi-wavelength follow-up observations of unidentified Fermi $\gamma$-ray sources \citep{Hessels2011, Ray2012, Braglia2020, Turchetta2024, Lu2025}. 

Here we present multi-wavelength follow up of two {\it Fermi}-LAT sources identified as spider candidates, for which radio/$\gamma$-ray pulsations have not been reported yet. We study their optical and near-infrared (NIR) emission, dominated by the companion star, thanks to new optical observations obtained in 2023-2024 with small robotic telescopes. Spiders present two distinct characteristic modulations in the optical-NIR light curves, either double-peaked with amplitudes of $\simeq\!0.3~\mathrm{mag}$, resulting from ellipsoidal variations associated with a tidally distorted companion star, or single-peaked with larger amplitudes of $\approx\! 1 ~\mathrm{mag}$, due to the irradiating pulsar wind heating the side of the companion facing the pulsar \citep{Breton2013, Turchetta2023}. The optical light curve can further reveal various properties of the companion, such as its effective temperature across the orbit and constrain the parameters of the binary system through light curve modeling \citep[e.g.,][]{Cho2018, Swihart2022, Sen2024}.

2FGL~J0846.0+2820 (J0846 hereafter) is a huntsman candidate, classified as such due to its long $\simeq\! 8.1~\mathrm{d}$ orbital period and giant secondary \citep{Swihart2017}. Optical spectroscopy revealed an eccentric ($e = 0.06$) $195 ~\mathrm{h}$ orbit, with a proposed $\simeq\! 0.8 ~\mathrm{M}_\odot$ partially stripped giant secondary and an inferred massive $\simeq\! 2.0 ~\mathrm{M}_\odot$ NS primary \citep{Swihart2017}. First detected as a $\gamma$-ray source in the {\it Fermi}-LAT 2FGL catalog \citep{Nolan2012}, the $\gamma$-ray flux dropped substantially in mid-2009, after which it was not included in the 3FGL and 4FGL catalogs. \citet{Swihart2017} reported an optical counterpart from the Catalina Sky Survey\footnote{\url{http://nesssi.cacr.caltech.edu/DataRelease/}} (CSS~J084621.9+280840, \citealt{Drake2009, Drake2014}) and a NIR counterpart from the Two Micron All Sky Survey of Point Sources\footnote{\url{https://irsa.ipac.caltech.edu/cgi-bin/Gator/nph-scan?mission=irsa&submit=Select&projshort=2MASS}} (2MASS~08462187+2808408, \citealt{Skrutskie2003, Skrutskie2006}) to J0846. No searches for a radio MSP in J0846 have been reported to date, to the best of our knowledge.

4FGL~J0935.3+0901 (J0935 hereafter) is a spider pulsar candidate. Optical spectroscopy of this system revealed double-peaked hydrogen and helium emission lines \citep{Wang2020} --- features commonly observed in transitional MSPs during the subluminous disc state \citep{Shahbaz_2019, Zelati_2014} --- suggesting that J0935 may belong to this class. A faint X-ray counterpart was also reported by \citet{Wang2020}. J0935 exhibits optical orbital modulations at a period of $\simeq\! 2.4 ~\mathrm{h}$, with no current companion mass inferences. Based on this short orbital period, with weak orbital modulation explained by a small inclination angle, this system has also been suggested to harbor a BW \citep{Halpern2022}. No radio pulsations were detected with the Five-hundred-meter Aperture Spherical radio Telescope (FAST) in the $1.05\!-\!1.45 ~\mathrm{GHz}$ band, which covered only 14\% of the orbit (\citealt{Zheng2022}; see also \citealt{Corcoran2023}).

\section{Observations and data analysis} \label{sec:observations_and_data_analysis}

\subsection{TJO Observations of 2FGL~J0846.0+2820} \label{sec:TJO_observations_of_J0846}
We obtained 1113 BVRI images of J0846 with the MEIA3 optical imaging camera on the 0.8-m Joan Oró Telescope (TJO\footnote{\url{https://montsec.ieec.cat/en/astronomy/tjo/}}) at the Montsec Observatory (OdM), spanning two observational periods: 527 images were taken over 27 observation nights in January-March 2023, and 586 images were taken over 17 observation nights in January-May 2024, ensuring a nearly full orbital phase coverage across the TJO 2023-2024 dataset. In Table~\ref{tab:setup}, we report details on the instrumental setup. The relatively bright magnitude of the target ($V\simeq16$~mag; \citealt{Swihart2017}) allowed for 60s exposures. All images were pre-processed (bias, dark and flat corrections) as part of TJO's standard calibration procedures. 

\subsection{LCO Observations of 4FGL~J0935.3+0901} \label{sec:LCO_observations_of_J0935}
We obtained 60 images of J0935 between the 14th and 28th of December, 2023, across two sites --- the Siding Spring Observatory and the McDonald Observatory --- using the Sinistro\footnote{\url{https://lco.global/observatory/instruments/sinistro/}} camera mounted on the 1-m telescope of the Las Cumbres Observatory (LCO\footnote{\url{https://lco.global/observatory/telescopes/1m0/}}), alternating $g'$, $r'$ and $i'$ filters. The observational setup used for J0935 is reported in Table~\ref{tab:setup}. For both J0846 and J0935, we also report the number of images that were left for the analysis after excluding images which were unsuitable for precise photometry (blurry or with marginally detected targets).

\begin{table}[ht!]
    \centering
    \caption{Comparison of observational setups for J0846 and J0935.}
    \begin{footnotesize}
    \begin{tabular}{lrr}
    \hline\hline
        \textbf{Parameter} & \textbf{J0846} & \textbf{J0935} \\
        \hline
        Obs. Period\footnote{Evening dates of first and last observation} & 2023: 01/20 -- 03/07 & 2023: 12/14 -- 12/28 \\
        & 2024: 01/28 -- 05/01 & \\
        Telescope & TJO-0.8m & LCO Siding Spring-1m \\
        & & LCO McDonald-1m \\
        Camera & MEIA3 & SINISTRO \\
        Bands & $BVRI$ & $g'r'i'$ \\
        Images\footnote{Number of images aquired (number of images suitable for precise photometry)} & 1113 (1069) & 60 (49) \\
        Magnitude\footnote{As given in literature of the respective target} & $15.4\!-\!16.7$\footnote{\citet{Swihart2017}} (BVR) & $20\!-\!21.5$\footnote{\citet{Wang2020}} (riz$_\mathrm{PS1}$) \\
        Exp. time (s) & $60$ & $300$ \\
        Airmass & $1.03\!-\!1.99$ & $1.08\!-\!1.58$ \\
        Pipeline\footnote{Image reduction pipeline applied before analysis} & ICAT\footnote{\citet{Colome09}} & BANZAI\footnote{\citet{McCully2018}} \\
        \hline
    \end{tabular}
    \end{footnotesize}
    \label{tab:setup}
\end{table}

\subsection{Differential Photometry} \label{sec:differential_photometry}
We performed differential photometry to measure our targets' magnitudes using an ensemble of comparison stars in the field of view (see, e.g., \citealt{Honeycutt1992} and \citealt{Warner2016}), thereby optimizing the precision of our target's photometry. We ensure stability of the ensemble of reference stars by enforcing a variability filter, quantified by the root-mean-square (RMS) deviation: we select reference stars with RMS deviations lower than $0.03$~mag, that are brighter than our targets and have similar colors. Calibrated reference magnitudes were obtained from the Pan-STARRS~1 catalog (PS1~DR1\footnote{\url{https://catalogs.mast.stsci.edu/panstarrs}}; \citealt{Chambers2019, PS1DR1}). Conversion from the PS1 griz magnitudes to the SDSS g'r'i'z' and Johnson-Cousins BVRI magnitudes was performed using a linear transformation (coefficients from Table~6 of \citealt{Tonry2012}).

For J0846, we used 12 stable reference stars ($1\!-\!2$ magnitudes brighter than the $V \simeq 16$~mag optical counterpart). We used the Source Extraction and Photometry (SEP\footnote{\url{https://sep.readthedocs.io/en/stable/}}) Python package (\citealt{Barbary2016}, based on the \textsc{SExtractor} software by \citealt{Bertin1996}), to identify all the sources in the field, determine their FWHMs, set the aperture radii and extract their counts. Because the TJO pointings were not stable, we used SEP to track our target and reference stars using their equatorial coordinates. An initial source extraction was performed to find the FWHMs of our reference stars; then we scaled the aperture radii to the mean reference FWHM in the second iteration. A scale factor of 0.8 was found to be optimal, minimizing sky contribution while still enclosing most source counts.

For J0935, we selected 5 stable reference stars $3\!-\!4$ magnitudes brighter than the reported $20\!-\!21.5 ~\mathrm{mag}$ of the target \citep{Wang2020}. The ULTRACAM\footnote{\url{https://cygnus.astro.warwick.ac.uk/phsaap/software/ultracam/html/index.html}} pipeline \citep{Dhillon07} was used to perform variable aperture photometry, measuring the flux within a circular region of J0935 and subtracting sky background using an annulus around the target aperture. We also scaled the source, inner- and outer-background radii to the seeing or FWHM, using scale factors of 1.5, 2.5 and 3.0, respectively.

\subsection{Phase Folding} \label{sec:phase_folding}
We applied barycentric correction to the recorded mid-exposure times (MJD-UTC), using the \texttt{utc\_tdb} function from the \texttt{barycorrpy} Python library \citep{Kanodia2018}. We phase folded our magnitudes at the orbital period ($P_\mathrm{orb}$), using as reference epoch ($T_0$) the time of inferior conjunction of the companion.

For J0846, \citet{Swihart2017} found a low but non-zero eccentricity ($e = 0.061 \pm 0.017$), $P_\mathrm{orb} = 8.13284 \pm 0.00043$~d and an epoch of periastron passage of the companion $T_\mathrm{p} = 2457007.9589^{+0.2378}_{-0.2907}$~BJD, from the radial velocity curve of the companion. We converted $T_\mathrm{p}$ to an epoch of inferior conjunction $T_0 = 57003.6^{+0.3}_{-0.4}$~MJD-TDB, assuming a quasi-circular orbit and using their reported argument of periastron, $\omega = 81^\circ.8^{+10.7}_{-12.9}$. Because the reported eccentricity is marginally significant, we also refitted the barycentric radial velocities measured by \citet{Swihart2017} assuming a circular orbit and found $T_0 = 57003.589 \pm 0.018$~MJD-TDB and $P_\mathrm{orb} = 8.13346 \pm 0.00037$~d, consistent with the previous measurements. We proceed with these latter orbital ephemeris to phase fold our photometric data, given the lower statistical uncertainties. The propagated error on the orbital phase was $\leq\! 0.017$ in 2023 and $\leq\! 0.020$ in 2024.

For J0935, we used the ephemeris reported by \citet{Halpern2022}: $P_\mathrm{orb}=0.10153276(36)$~d and a time of ascending node of the putative pulsar $T_\mathrm{asc}=59584.4015(7)$~MJD-TDB. We converted the latter to an epoch of companion inferior conjunction $T_0=59584.4269(7)$~MJD-TDB. The orbital phase error was estimated to be $0.026$ in our observed epoch, from the error propagation of $P_{\mathrm{orb}}$ and $T_0$. 

\subsection{{\it Fermi}-LAT $\gamma$-Ray Analysis} \label{sec:Fermi_LAT_gamma-ray_analysis}
To investigate the long-term (2008-2025) $\gamma$-ray variability of J0846, we extracted a light curve using all {\it Fermi}-LAT data available as of August 2025. We selected $0.1\!-\!100$ GeV \texttt{SOURCE} class photons within a $15^\circ$ radius region of interest (RoI), with zenith angles $\leq\!90^\circ$. We included photons detected between 2008 August~4 and 2025 August~14, using the \texttt{P8R3\_SOURCE\_V3} instrument response functions (IRFs). The data were divided into 34 time bins (the first 33 bins are 180~d long and the last one is 278~d). For each bin, we performed a binned likelihood analysis using \texttt{fermipy} \citep{Wood2017}. The sky model was based on the 4FGL-DR4 catalog \citep{Abdollahi20, Abdollahi22, Ballet23}, together with the Galactic and isotropic diffuse background models \texttt{gll\_iem\_v07.fits}, \texttt{iso\_P8R3\_SOURCE\_V3\_v1.txt}. Since J0846 is not included in 4FGL, we added a point source at its position (RA=$131.511^\circ$, DEC=$28.348^\circ$), modeled with a simple power-law with the photon index fixed at $\Gamma=2.7$, following \citet{Swihart2017}. We allowed the normalizations of J0846, all sources within $6^\circ$, and the diffuse backgrounds to vary during the fit. 

\subsection{Chandra X-Ray Analysis} \label{sec:chandra}
We analyzed the only {\it Chandra} observation of J0846 available at the time of writing, taken with the ACIS-I detector on 2017 June~3, for a total net exposure of 19.7~ksec (2.8\% of the orbit; OBSID: 17863, PI: Strader). A source coincident with the optical location of J0846 is detected with about 10 net counts in the $0.3\!-\!10$~keV band (which corresponds to the cataloged source 2CXO~J084621.8+280840, also identified as an X-ray counterpart to J0846 by \citealt{Koljonen2025}). We extracted source and background spectra using \textsc{CIAO} (v.~4.10, \citealt{CIAO2026}) and circular regions of $2''$ and $21''$ radius, respectively.

We fitted the resulting net spectrum in the $0.3\!-\!10$~keV band within \textsc{XSpec} (v.~12.15.1, \citealt{XSpec}). Due to the low number of source counts, we fixed the absorbing hydrogen column density to the total value along the line of sight \citep[$n_\mathrm{H}$=4.5$\times$10$^{20}$~cm$^{-2}$; from][]{HI4PI} and fitted the spectrum using the C statistic\footnote{\url{https://heasarc.gsfc.nasa.gov/docs/software/xspec/manual/node119.html}}.

\section{Results} \label{sec:results}

\subsection{A State Change from the Optical Light Curves of 2FGL~J0846.0+2820} \label{sec:results_J0846}
\begin{figure*}[h!]
    \centering
    \plotone{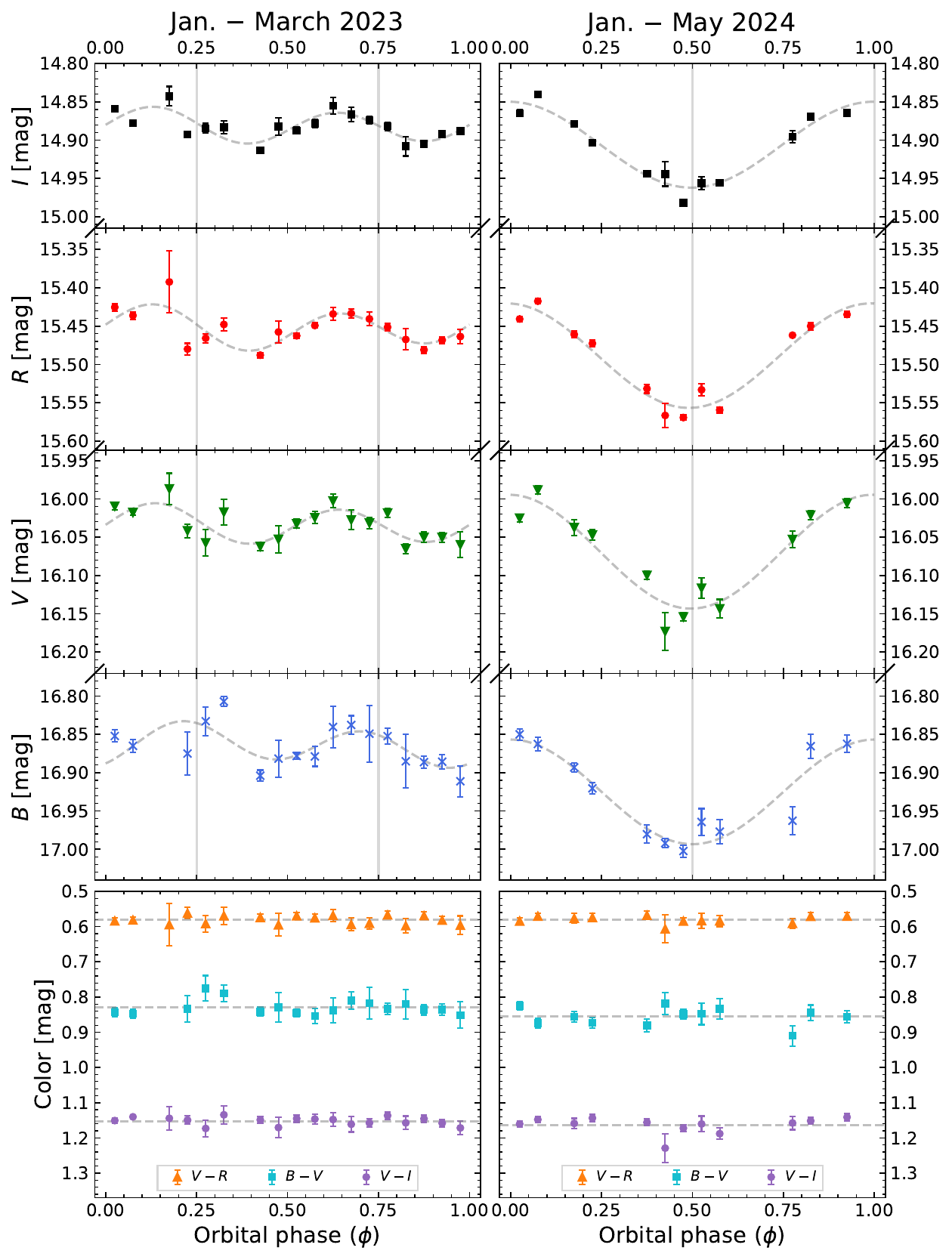}
    \caption{Folded optical-NIR light curves of J0846 (where phase 0 corresponds to inferior conjunction of the companion, see Sec.~\ref{sec:phase_folding}). \textit{Top panels:} Apparent BVRI magnitudes, plotted separately for our 2023 (left) and 2024 (right) observations. The data are binned with phase intervals of 0.05, and fitted with a second order (2023) and first order (2024) Fourier series. \textit{Bottom panels:} Binned apparent $B\!-\!V$, $V\!-\!R$ and $V\!-\!I$ colors and mean values.}
    \label{fig:J0846_LC_Col}
\end{figure*}

The average apparent magnitudes in our 2023-2024 TJO observations of J0846 are $B = 16.89~\mathrm{mag}$, $V = 16.05~\mathrm{mag}$, $R = 15.47~\mathrm{mag}$ and $I = 14.89~\mathrm{mag}$ \citep[about $\sim\!0.1$~mag fainter than the averages in][]{Swihart2017}. The corresponding standard deviations of the magnitudes in each filter are 0.06, 0.05, 0.05 and 0.04~mag, respectively. Our long-term optical-NIR light curves reveal a larger spread in apparent BVRI magnitudes in 2024 (peak-to-peak amplitudes of $0.2\!-\!0.4$~mag) compared to 2023 (peak-to-peak amplitudes of $0.1\!-\!0.2$~mag), suggesting a difference in orbital modulations. We therefore choose to phase-fold our data separately for the two epochs. We also re-bin our data in intervals of 0.05 orbital phases to show the modulation patterns more clearly. The resulting binned orbital light curves of J0846 are shown in Figure~\ref{fig:J0846_LC_Col}. The apparent $B\!-\!V$, $V\!-\!R$ and $V\!-\!I$ colors are calculated as the difference between the phase-folded binned BVRI magnitudes at each orbital interval, and are shown in the bottom panels of Figure~\ref{fig:J0846_LC_Col}.

We find a state change in the folded optical-NIR light curves of J0846, when comparing observations taken one year apart: from double-peaked in 2023 to single-peaked in 2024. To quantify and highlight this state change, we fitted a 2nd (2023) and 1st order (2024) Fourier series to the orbital light curves and used these to estimate the orbital phase of the minima/maxima. The binned 2023 BVRI light curves have peak-to-peak amplitudes of $\sim\!0.07\!-\!0.10~\mathrm{mag}$. In 2023 (VRI bands), we find two flux minima at phases $\simeq\!0.4$ and $\simeq\!0.9$, and two maxima at phases $\simeq\!0.15$ and $\simeq\!0.65$. The B-band light curve is more noisy, but seems delayed or shifted by about 0.05 in orbital phase. The 2024 data have binned peak-to-peak amplitudes of $\sim\!0.14\!-\!0.18~\mathrm{mag}$, displaying near-equal flux maxima but dimmer flux minima than in 2023. In 2024, the flux minimum occurs at phase $\simeq\!0.5$, while the flux maximum occurs at phase $\simeq\!0.0$.

The $B\!-\!V$, $V\!-\!R$ and $V\!-\!I$ color curves are all nearly flat in 2023 and 2024, with no clear variable trends and RMS values ranging from 0.010 to 0.025~mag. This indicates nearly constant colors along the orbit, corresponding to little or no temperature changes across the companion star of J0846, and thereby weak to absent irradiation (see discussion in Sec.~\ref{sec:irradiation_J0846}).

\subsection{An Irradiated Single-Peaked Orbital Light Curve in 4FGL~J0935.3+0901} \label{sec:results_J0935}
\begin{figure*}[ht!]
    \centering
    \plotone{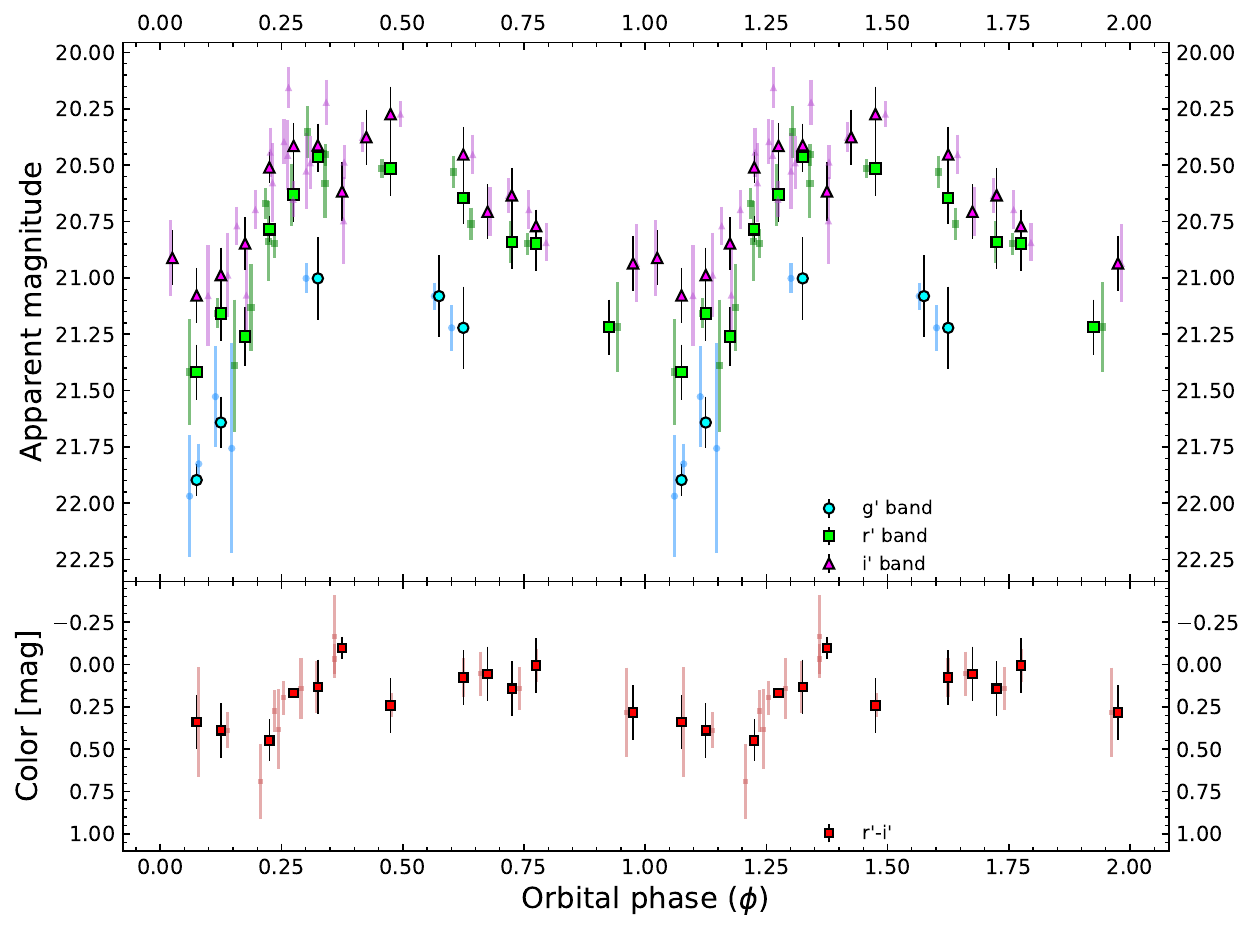}
    \caption{Folded optical light curves of J0935 (where phase 0 corresponds to inferior conjunction of the companion, see Sec.~\ref{sec:phase_folding}). Small/shaded symbols show the original data points, and large/bright symbols show binned data. Two orbital cycles are shown for display purposes. \textit{Top panel:} Apparent magnitudes in the $g'$, $r'$ and $i'$ bands. \textit{Bottom panel:} Observed $r'\!-\!i'$ color index.}
    \label{fig:J0935_LC_Col}
\end{figure*}

We show in Figure~\ref{fig:J0935_LC_Col} the folded orbital $g'$, $r'$ and $i'$-band light curves of J0935. We show the original data points, and re-bin them in 0.05 phase bins to display the variable trend more clearly. We find an average $r'$-band magnitude of $20.8~\mathrm{mag}$, consistent with that reported by \citet{Wang2020} and \citet{Halpern2022}, but our data have full multi-band coverage of the orbit. We also find the mean $g' = 21.5~\mathrm{mag}$ and $i' = 20.6~\mathrm{mag}$, as well as standard deviations of 0.4, 0.3 and 0.3~mag in the $g'$, $r'$ and $i'$ bands, respectively. All three light curves exhibit consistent minima at orbital phase $\simeq\! 0.06$. The light curves show a single peak per orbit around phase $\simeq\! 0.5$, and peak-to-peak amplitudes of $1.0\pm0.3~\mathrm{mag}$, $1.1\pm0.3~\mathrm{mag}$ and $0.9\pm0.3~\mathrm{mag}$ in $g'$, $r'$ and $i'$, respectively.

The $r'\!-\!i'$ color curve also indicates a broad maximum at orbital phase $\simeq0.5$, when the companion is at superior conjunction, whereas the minimum occurs around phase $0.1$ (similarly to what was observed in the orbital light curves). A 1st order Fourier series fit to the data reveals that the $r'\!-\!i'$ color changes by about 0.4~mag, suggesting a possible variation of the companion's temperature between superior and inferior conjunctions of the companion (see Sec.~\ref{sec:discussion_J0935} for further discussion). This, together and correlated with the single-peaked light curves, is characteristic of a system dominated by irradiation \citep[see, e.g.,][and references therein]{Turchetta2023}.

\subsection{High-Energy Emission from J0846} \label{sec:gamma-ray_variability}
\citet{Swihart2017} reported that the $\gamma$-ray flux of J0846 dropped in mid-2009, corresponding with an increased variation in the optical brightness. Since then, the source has not been detected by {\it Fermi}-LAT. To investigate possible changes in the $\gamma$-ray flux of J0846 during 2023 and 2024, when the optical state change occurred, we constructed a long-term {\it Fermi}-LAT light curve as detailed in Section~\ref{sec:Fermi_LAT_gamma-ray_analysis}.

\begin{figure}[ht!]
    \centering
    \includegraphics[width=\columnwidth]{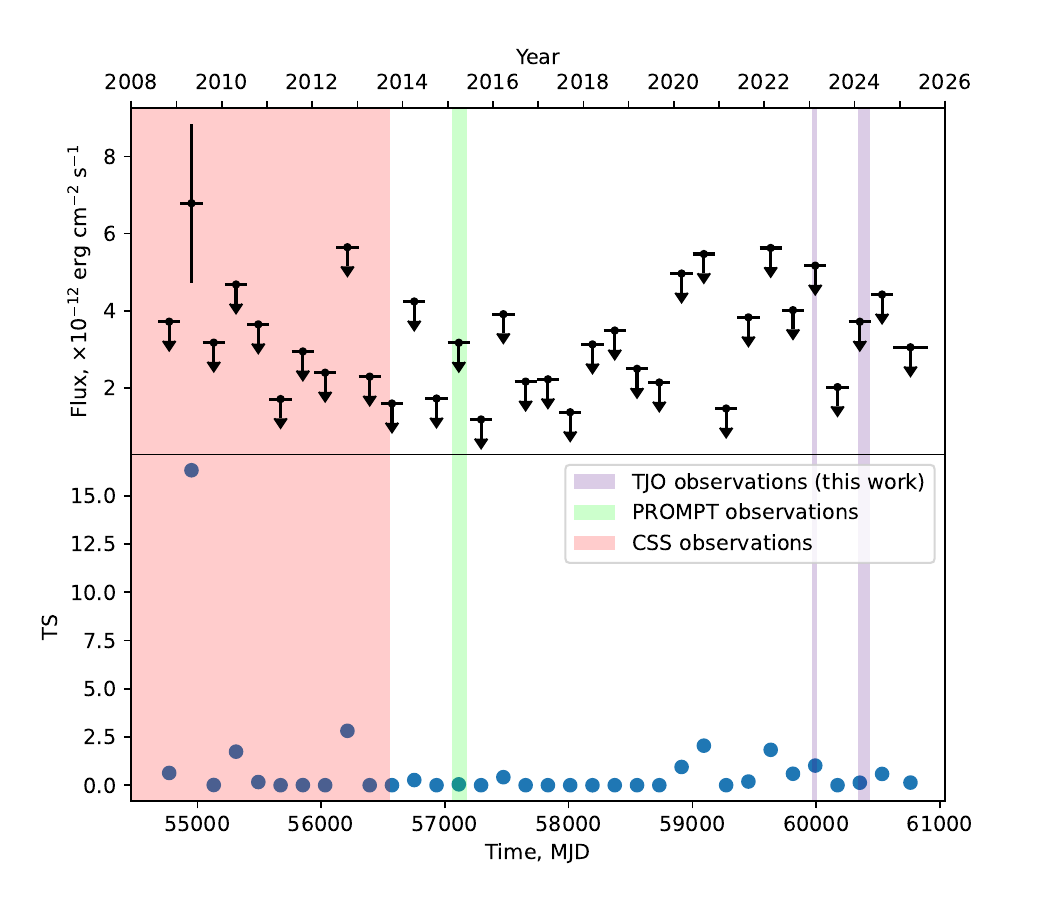}
    \caption{Long-term {\it Fermi}-LAT light curves of J0846. \textit{Top panel:} Photon flux plotted over time. \textit{Bottom panel:} Test statistics of J0846 over time. Shaded purple regions show the dates of the TJO observations presented in this work. Shaded red and green regions show the dates of CSS and PROMPT observations, respectively, presented in \citealt{Swihart2017}.}
    \label{fig:J0846_LAT_LC}
\end{figure}

The resulting light curve is shown in Figure~\ref{fig:J0846_LAT_LC}. Overall, the light curve is consistent with that reported by \citet{Swihart2017}. Some differences in the TS, flux values, or upper limits may arise from the updated IRFs and background models. The source is significantly detected (with test statistics TS$>\!9$) only in the second bin (2009 January 31 to 2009 July 30) with the $0.1\!-\!100$~GeV flux of $(17.6\pm5.4)\times10^{-9}$ photons~cm$^{-2}$~s$^{-1}$, while for all other bins --- including those covering 2023 and 2024 --- we place upper limits on the flux. Upper limits range between $(3\!-\!5) \times10^{-9}$~photons~cm$^{-2}$~s$^{-1}$, with the scatter driven by variability of nearby sources in the RoI. Therefore, we find no evidence for significant changes in the $\gamma$-ray flux associated with the optical state change of J0846, and confirm that the source was only detected in that February-July 2009 period. The $0.1\!-\!100$~GeV energy flux during that 6-month period was $F_\gamma = (6.8\pm2.1)\times10^{-12}$~erg~s$^{-1}$~cm$^{-2}$, which corresponds to a luminosity $L_\gamma \simeq 1.2\times10^{34}$~erg~s$^{-1}$ at 3.76~kpc (from \citealt{Koljonen2023}; see Sec.~\ref{sec:sed_J0846} for discussion).

We also analyzed the X-ray counterpart to J0846 detected with {\it Chandra} in 2017, as explained in Section~\ref{sec:chandra}. The X-ray spectrum can be fitted with a simple absorbed power law model (\textsc{tbabs*powerlaw}), yielding a photon index in the $2.6\!-\!4.9$ range (68\% confidence level, assuming $n_\mathrm{H}$=4.5$\times$10$^{20}$~cm$^{-2}$). From the best fit, we estimate a $0.5\!-\!10$~keV energy flux $F_\mathrm{X}\simeq1.1\times10^{-14}$~erg~s~cm$^{-2}$ (consistent with 2CXO catalog values), which translates to an X-ray luminosity of $L_\mathrm{X}\simeq1.8\times10^{31}$~erg~s$^{-1}$ at 3.76~kpc (from \citealt{Koljonen2023}; see Sec.~\ref{sec:sed_J0846} for discussion).

\section{Discussion} \label{sec:discussion}

\subsection{2FGL~J0846.0+2820: An Active Giant Secondary Under-Filling its Roche Lobe?}
\label{sec:discussion_J0846}

\subsubsection{Lack of Ellipsoidal Modulation} 
\label{sec:ellipsoidal_J0846}
The double-peaked, low-amplitude 2023 light curves of J0846 (Fig.~\ref{fig:J0846_LC_Col}, left), in combination with the flat color curves, are reminiscent of a system dominated by ellipsoidal modulation, which can arise when the companion star is close to filling its Roche lobe and tidally distorted. However, the 2023 flux maxima are not in orbital quadratures ($\phi=0.25, 0.75$; when the flux and projected area from a tidally distorted star would be maximum). As shown in Figure~\ref{fig:J0846_LC_Col} (left), we find these flux maxima occur earlier in the orbit, shifted by $\sim\!0.1$ orbital phases (more than 5 times the propagated uncertainty in $\phi$, cf. Sec.~\ref{sec:phase_folding}). Furthermore, the V-band peak-to-peak amplitude of J0846 in 2023 was only $\Delta V\lesssim 0.1$~mag, while the ellipsoidal light curves from the two confirmed huntsmen and most RBs show higher $\Delta V\gtrsim 0.2-0.3$~mag \citep[e.g.,][]{Strader2015,Linares17,Strader2025}.

Thus, our 2023 optical data of J0846 are not consistent with a tidally-distorted (nearly) Roche lobe-filling companion. Since the orbital period is robust \citep[][from the radial velocity curve]{Swihart2017}, we conclude that the companion or secondary star in J0846 is severely under-filling its Roche lobe, so that ellipsoidal modulation is not detectable. This also casts doubt on binary parameters inferred from light curve fits in this system \citep[orbital inclination, Roche lobe filling factor; ][]{Swihart2017}, since they relied mostly on modeling the ellipsoidal signal (which we argue is absent).

Alternatively, a low orbital inclination could result in a low-amplitude ellipsoidal modulation, but then we would expect this modulation to be persistently low and constant in orbital phase. To verify this, we compiled the CSS light curves of the optical counterpart to J0846 (CSS~J084621.9+280840), including the 2005-2013 CSS data presented by \citet{Swihart2017}, along with 216 additional data points between 2013 and 2016 (CSDR3). We find a gradual long-term increase in the CSS V-band flux from April 2005 to April 2016, similar to the --0.011~mag~yr$^{-1}$ reported by \citet{Swihart2017}. Folding the light curves from 12 separate segments, we find variable amplitudes, different light curve shapes and varying phases of flux maxima and minima. This is not consistent with a low-amplitude ellipsoidal modulation.

\begin{figure}[ht!]
    \centering
    \includegraphics[width=\columnwidth]{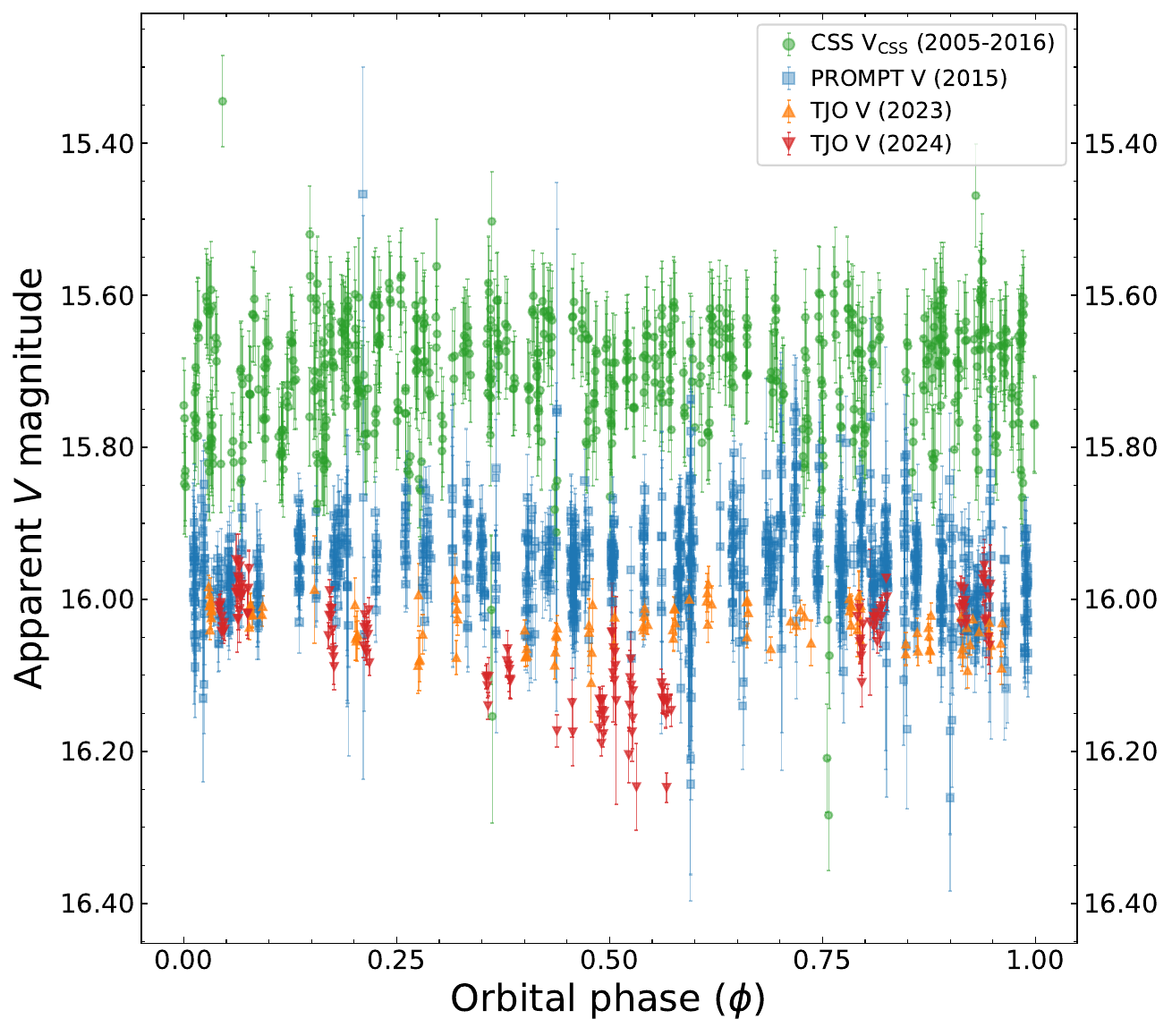}
    \caption{Folded light curves of J0846 in the V band, from observations made with the CSS \citep{Drake2009}, PROMPT \citep{Swihart2017} and TJO (this work).}
    \label{fig:J0846_LC_V}
\end{figure}

We show in Figure~\ref{fig:J0846_LC_V} the folded orbital light curves from J0846, using all the V-band magnitudes measured between 2005 and 2024 (from CSS, PROMPT and the TJO observations presented in this work). The PROMPT V-band magnitudes are retrieved from the machine-readable version of Table 1 of \citet{Swihart2017}.  Note that the CSS V-band equivalent differs from the Johnson-Cousins V filter used in the PROMPT and TJO observations, so the CSS-PROMPT/TJO magnitude shift in Figure~\ref{fig:J0846_LC_V} is partly instrumental. Regardless, there is no consistent orbital modulation and the long-term CSS folded light curves show a similar scatter ($\simeq\! 0.2$~mag) across all orbital phases. In summary, the optical variability from J0846 is different than that seen in confirmed spider MSPs, and does not seem correlated with the orbital phase. In the absence of irradiation (Sec.~\ref{sec:irradiation_J0846}), ellipsoidal modulation cannot appear and disappear on time scales of years (since the filling factor cannot change drastically on those time scales). We then conclude that the Roche lobe filling factor in J0846 is low, so that the binary is detached and other mechanisms dominate the optical variability (briefly discussed at the end of Sec.~\ref{sec:irradiation_J0846}).

\subsubsection{Lack of Irradiation} 
\label{sec:irradiation_J0846}
The single-peaked, higher-amplitude 2024 light curves of J0846 (Fig.~\ref{fig:J0846_LC_Col}, right) are also difficult to interpret within the spider MSP identification, as the flux minimum occurs around the companion's superior conjunction with no apparent color change. A transition from an ellipsoidal- to an irradiation-dominated regime was observed in the ``face-changing" companion of the RB PSR~J1048+2339 \citep{Yap2019}: the minimum flux at inferior conjunction of the companion ($\phi=0$) remained constant while the flux at superior conjunction ($\phi=0.5$) increased, turning a double-peaked ellipsoidal light curve into a single-peaked irradiation case. Instead, we find that the flux maximum in J0846 remains approximately constant (about 16.0~mag in V), but shifts in phase between 2023 and 2024, while the flux minimum decreases. Most importantly, the flux maximum in the 2024 J0846 observations occurs at inferior conjunction of the companion/secondary ($\phi=0$), the exact opposite of what we see in strongly irradiated spiders \citep[e.g.,][]{Breton2013}. As mentioned in Section~\ref{sec:results_J0846}, no irradiation pattern is detectable in the 2024 color curves either (Fig.~\ref{fig:J0846_LC_Col}, bottom right panel).

We also estimate the effective temperature of J0846 from its intrinsic color indices. De-reddening is performed using the best-fit SDSS color excess $E(g\!-\!r) = 0.054~\mathrm{mag}$ from the Bayestar19 3D dust maps \citep[][for a distance $d>0.2$~kpc]{Green2019}, converted to the corresponding BVRI extinction coefficients following \citet[][their Table 6]{Schlafly2011}, assuming an extinction law with $R_V = 3.1$. We find the mean intrinsic (de-reddened) colors $B\!-\!V = 0.80 \pm 0.06~\mathrm{mag}$, $V\!-\!R = 0.55 \pm 0.04~\mathrm{mag}$ and $V\!-\!I = 1.10 \pm 0.04~\mathrm{mag}$, which are consistent with a companion star with an effective temperature of $4830\pm270~\mathrm{K}$ in both 2023 and 2024 \citep{Pecaut2013}.

We can now quantify the irradiation of the companion of J0846 by computing the ratio between the putative pulsar wind flux intercepted by the companion star and the companion intrinsic flux, $f_\mathrm{sd}$, introduced and defined by \citet{Turchetta2023} as
\begin{equation}\label{eq:f_sd}
    f_\mathrm{sd} \equiv \frac{L_\mathrm{sd}}{L_\mathrm{2}}\frac{R_2^2}{a^2},
\end{equation}
for a spin-down luminosity $L_\mathrm{sd}$, a companion intrinsic luminosity $L_\mathrm{2}$, a radius $R_2$ and an orbital separation $a$. We obtain $f_\mathrm{sd}\simeq 0.0091 \pm 0.0025 \left[ L_\mathrm{sd} / 10^{34}\mathrm{~ erg~s}^{-1} \right]$ from their Equation~(3), using our $T_2 = 4830 \pm 270$~K and $P_\mathrm{orb} = 195.203 \pm 0.009$~h \cite[as well as $M_1 + M_2 = 2.7 \pm 0.6 ~\mathrm{M}_\odot$ as inferred by][]{Swihart2017}. With no pulsar detection available, we can assume that the primary exhibits typical spin-down luminosities of $L_\mathrm{sd} \simeq 10^{33}-10^{35} \mathrm{~erg~s}^{-1}$ observed in spider systems \citep[e.g.,][]{Linares2021}. With this, we estimate that $f_\mathrm{sd}$  is in the $10^{-3}\!-\!10^{-1}$ range, if J0846 contains a typical MSP. Such a low flux ratio places J0846 well below the irradiation range, compared to the corresponding values from spiders \citep[][found the transition at $f_\mathrm{sd}\simeq2\!-\!4$]{Turchetta2023}. We therefore conclude that, given the orbital size and companion temperature, irradiation is negligible in J0846, and a change to an irradiation-dominated state in 2024 (or anytime) can be safely excluded.

Given the long-term variability (see Sec.~\ref{sec:ellipsoidal_J0846}), which does not seem correlated with the orbit, we attribute the state change of J0846 between 2023 and 2024 to variable stellar activity and starspots. Migrating dark spots can explain the variable amplitude and shifting phase of the light curve extrema, perhaps combined with asynchronous rotation. \citet{vanStaden2016} presented evidence for an active asynchronous companion in the RB PSR~J1723--2837, which has a relatively long $P_\mathrm{orb} \simeq 14.8$~hr, and this is presumably more common in wider-orbit systems like J0846 (with $P_\mathrm{orb} \simeq 195.2$~hr) since tidal interactions are weaker.

\subsubsection{Masses and Orbital Parameters} 
\label{sec:masses}
Let us now reassess the nature of the primary star in J0846, with mass $M_1$ and radius $R_1$ a priory unknown. In the standard definition, the primary star is more massive, so that the mass ratio $q\equiv M_2/M_1 < 1$. The semi-amplitude of the optical radial velocity curve ($K_2=54.4$~km~s$^{-1}$, as determined by \citealt{Swihart2017}) gives an absolute lower limit $M_1>P_\mathrm{orb} K_2^3 / 2\pi G=0.14~\mathrm{M}_\odot$ (where $G$ is the gravitational constant). The same authors measured the rotational broadening projected onto the line of sight ($v \sin i$), from a high-resolution ($R\sim 36000$) optical ($\sim 3900\!-\!8100$~\AA) spectrum of the companion. From that, they inferred a mass ratio $q=0.40\pm0.04$ using the relation from \citet{Wade88}:
\begin{equation}
    v \sin i \simeq 0.462~K_2~q^{1/3}~(1+q)^{2/3}.
\label{eq:vsini}
\end{equation}
We note, however, that this relation assumes a companion that is {\it both} tidally locked with the orbit and filling its Roche lobe. None of these assumptions are known to hold for J0846, and we have argued that the companion is far from filling its Roche lobe (Sec.~\ref{sec:ellipsoidal_J0846}). In that case, Equation~\ref{eq:vsini} gives a lower limit on $q$, so we can only conclude that $q>0.4$ if the star is tidally locked. In other words, the mass ratio $q$ can be well above 0.4, since we argued that the companion is severely under-filling its Roche lobe. We note that both confirmed huntsman pulsars had well constrained $q\simeq0.2$ \citep[0.171 and 0.182 in J1417 and J1947,][respectively]{Camilo2016,Strader2025}. But we have also suggested that the secondary star in J0846 may be rotating asynchronously; in that case $q$ is unconstrained.

The orbital inclination $i$ is also ill-constrained. \citet{Swihart2017} found $i=26^\circ-34^\circ$ from their ELC light curve fits. However, we have argued that ellipsoidal modulation is absent in J0846 (Sec.~\ref{sec:ellipsoidal_J0846}), which would invalidate these constraints. The low $i$ could be an artifact of trying to model a low-amplitude modulation, which would in turn bias the previous estimates towards high masses. We conclude that both $i$ and $q$ are ill-constrained with the currently available data, and thus the masses $M_1$ and $M_2$ are basically unknown ($K_2$ and $P_\mathrm{orb}$ with the reported $q=0.4$ and a median $i=60^\circ$ would imply $M_1\simeq0.41$~M$_\odot$ and $M_2\simeq0.16$~M$_\odot$, but we stress that these are not actual constraints).

\subsubsection{Broadband SED: A Double Giant Alternative} 
\label{sec:sed_J0846}
\begin{figure*}[ht!]
    \centering
    \epsscale{0.55}
    \plotone{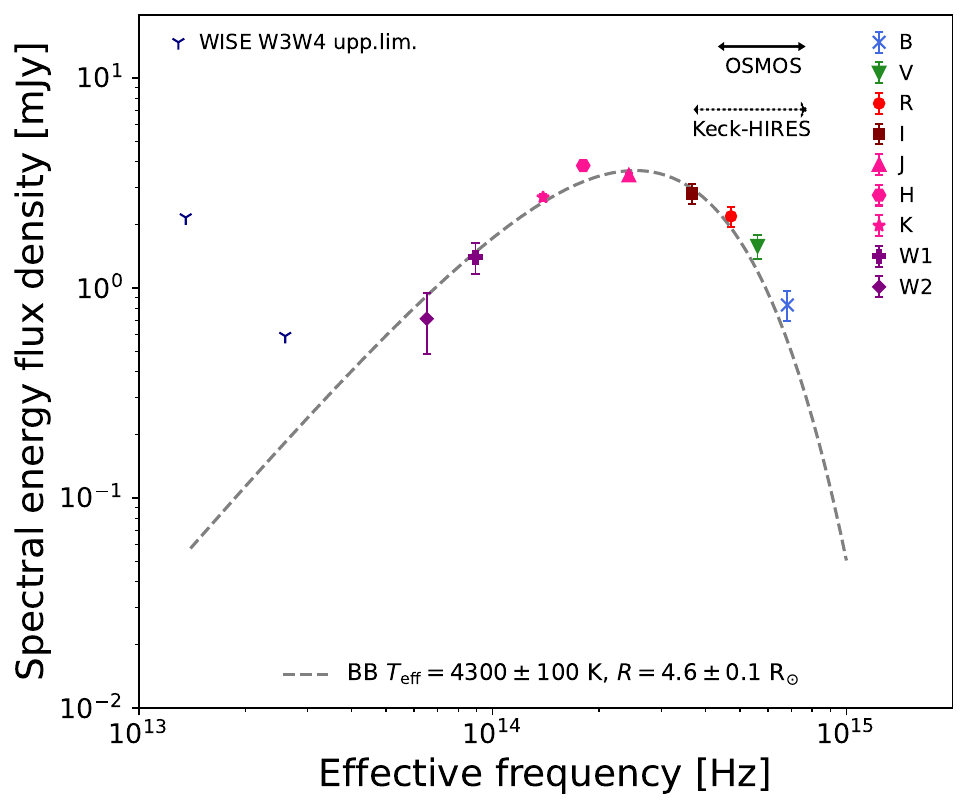}
    \plotone{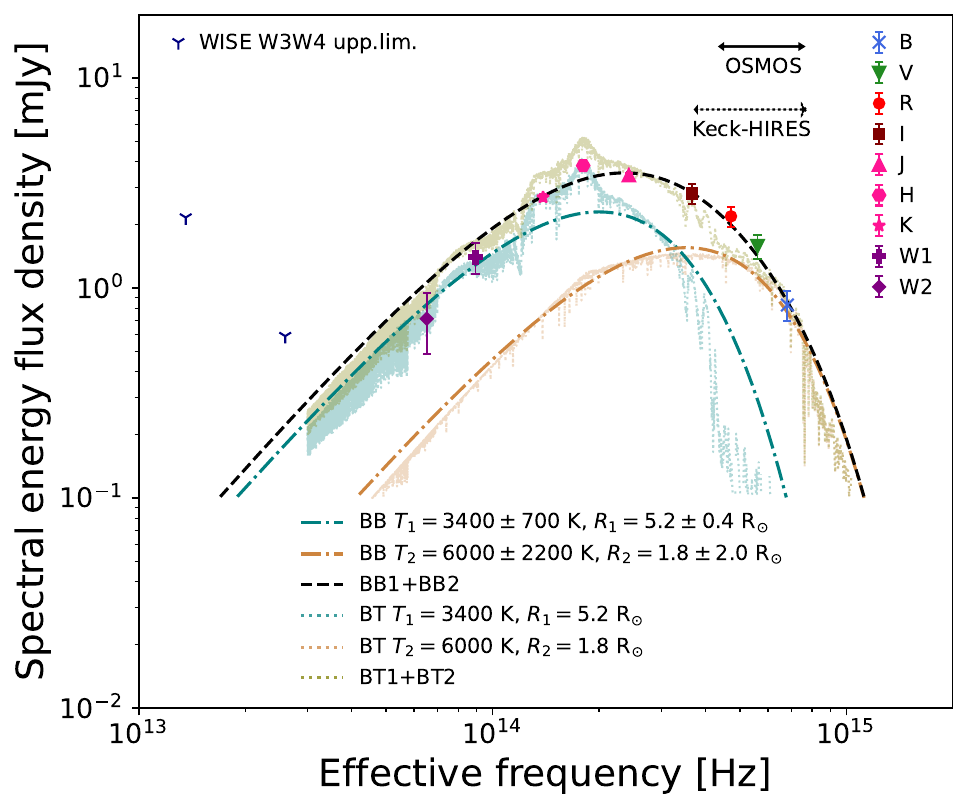}
    \caption{Spectral energy distribution of J0846 with blackbody fits. The ``OSMOS" and ``Keck-HIRES" arrows show the optical bands where previous spectroscopy was obtained \citep[3960-6840 Å and 3900-8100 Å;][]{Swihart2017}. {\it Left:} A single blackbody fit to the intrinsic TJO BVRI, 2MASS JHK and WISE W1W2 flux densities. {\it Right:} A 2-blackbody fit to the intrinsic TJO BVRI, 2MASS JHK and WISE W1W2 flux densities (black dashed line), composed of one cooler body 1 (cyan dash-dotted line) and one hotter body 2 (beige dash-dotted line).}
    \label{fig:J0846_SED}
\end{figure*}

To study the spectral energy distribution (SED) of J0846, we use the optical BVRI magnitudes (the average of our 2023-2024 TJO observations; Sec.~\ref{sec:TJO_observations_of_J0846}), the NIR JHK magnitudes from the 2MASS counterpart (2MASS~08462187+2808408, observed on 1998-02-03) and the IR WISE W1W2 magnitudes of J0846 retrieved from the ALLWISE catalog\footnote{\url{https://irsa.ipac.caltech.edu/data/WISE/docs/release/AllWISE/}} \citep{Wright2010}. We de-redden all magnitudes using $E(g\!-\!r) = 0.054~\mathrm{mag}$, as explained in Section~\ref{sec:irradiation_J0846}, and present the corresponding intrinsic flux densities in Figure~\ref{fig:J0846_SED}.

We fit the resulting SED with a simple blackbody model, fitting for effective temperature ($T$) and radius ($R$) and assuming a distance of $d=3.76$~kpc from the Gaia-DR3 measured parallax \citep{Gaia2023,Koljonen2023}. We note that this more recent distance measurement is not consistent with the previous estimate for J0846 \citep[$d\simeq7.0\!-\!8.3~\mathrm{kpc}$,][]{Swihart2017}. We fit our average flux density measurements using the \texttt{scipy.optimize.curve\_fit}\footnote{\url{https://docs.scipy.org/doc/scipy/reference/generated/scipy.optimize.curve_fit.html}} Python function \citep{2020SciPy}, which uses a non-linear least squares method with weighted standard deviation errors. We use an initial temperature guess of $4830~\mathrm{K}$ (as estimated in Sec.~\ref{sec:irradiation_J0846}) and an initial radius guess of $7.1~\mathrm{R}_\odot$ (as postulated by \citet{Swihart2017} for a Roche lobe filling companion). The uncertainties on the best-fit parameters are estimated from the square root of the diagonals of the co-variance matrix. The WISE W3 and W4 upper limits are shown in Figure~\ref{fig:J0846_SED}, but are not considered in our blackbody fitting.

First, we fit the full optical-IR (BVRIJHKW1W2) SED with a single blackbody, and obtain $T_\mathrm{eff}=4300 \pm 100~\mathrm{K}$ and $R=4.6 \pm 0.1~\mathrm{R}_\odot$. This fit, shown in the left panel of Figure~\ref{fig:J0846_SED}, does not capture the overall shape of the SED ($\chi^2/\mathrm{dof} = 34.2/7 = 4.9$), so we do not discuss it further. Then, we fit only the optical (BVRI) data points, and find an effective temperature of $T_2=5000 \pm 300~\mathrm{K}$ and a blackbody radius of $R_2=3.4 \pm 0.5~\mathrm{R}_\odot$ for the optical secondary in J0846 ($\chi^2/\mathrm{dof} = 0.3/2 = 0.1$). This temperature estimate agrees with the $5000\!-\!5250~\mathrm{K}$ range found by \citet{Swihart2017}, but we find a smaller $R_2$, driven by our smaller Gaia-DR3 $d$ value.

Interestingly, the NIR-IR (JHKW1W2) flux densities seem to trace a different component in the SED. Thus, we fit a two-blackbody model to the full BVRIJHKW1W2 SED. The resulting two-blackbody model is shown in Figure~\ref{fig:J0846_SED} (right), and provides a better fit to the broadband optical-IR SED ($\chi^2/\mathrm{dof} = 19.9/5 = 4.0$). This includes a hotter body dominating in the optical ($T_2 = 6000 \pm 2200 ~\mathrm{K}$, $R_2= 1.8 \pm 2.0~\mathrm{R}_\odot$) and a cooler body dominating the NIR-IR (with $T_1 = 3400 \pm 700 ~\mathrm{K}$ and $R_1= 5.2 \pm 0.4~\mathrm{R}_\odot$). We propose that this latter component is the primary star in J0846: a cooler red giant (unnoticed hitherto) which dominates in the NIR and IR bands, but is barely visible in the optical (where the radial velocities were measured: the OSMOS and Keck-HIRES spectral ranges are marked with arrows in Fig.~\ref{fig:J0846_SED}). We also plot two representative stellar atmosphere models from the BT-Settl library\footnote{\url{https://svo2.cab.inta-csic.es/theory/newov2/index.php}} \citep{Allard2011}, closely matching the blackbody $T_1 \text{, } R_1$ and $T_2 \text{, } R_2$ that we find, as well as their sum (see Figure~\ref{fig:J0846_SED}, right).

Let us now review the arguments for/against the spider MSP classification, in each wavelength range.
\begin{itemize}
    \item {\it Radio}. We searched for radio counterparts to J0846 within a $8''$ radius around its optical position, but found no matches in the 1.4 GHz \textit{NRAO VLA Sky Survey} \citep[NVSS;][]{1998AJ....115.1693C}, the 1.4 GHz \textit{Faint Images of the Radio Sky at Twenty Centimeters} survey \citep[FIRST;][]{2015ApJ...801...26H}, the 3 GHz \textit{Very Large Array Sky Survey} \citep[VLASS;][]{2021ApJS..255...30G}, or the 0.1--20 GHz Combined Radio Multi-Survey Catalog of Fermi Unassociated Sources compiled by \citet{2023ApJ...943...51B}. Thus, no radio source is known at the optical location of J0846, neither pulsed nor continuum.

    \item {\it Infrared}. The 2MASS NIR and WISE IR magnitudes suggest a different component, as explained above (Fig.~\ref{fig:J0846_SED}, right). The TJO and 2MASS observations are not simultaneous, which could affect this SED comparison if the variability is strong. Extrapolating the long-term trend of --0.011~mag~yr$^{-1}$ found from the 2005-2013 optical CSS magnitudes by \citet[][their Figure~7]{Swihart2017}, we estimate that the system could have been 0.28~mag {\it fainter} in 1998 than in 2023. But in the 2MASS 1998 SED, the K band is about 0.6~mag  {\it brighter} than the extrapolation of the hotter secondary to low frequencies, so this explanation seems unlikely.

    \item {\it Optical}. The optical variability of J0846 makes it different than any confirmed spider. We have argued in Sections~\ref{sec:ellipsoidal_J0846} and \ref{sec:irradiation_J0846} that it shows no signs of interaction (tidal or irradiation), which is a ubiquitous feature of spider MSPs. Furthermore, the measured $K_2$ is quite low (54~km~s$^{-1}$) compared to the two confirmed huntsman pulsars \citep[with 80 and 116~km~s$^{-1}$;][]{Strader2015,Strader2025}, which suggests a lower $M_1$ in J0846, if $i$ is similar.

    \item {\it X-rays}. The X-ray source described in Section~\ref{sec:gamma-ray_variability} has a luminosity ($L_\mathrm{X}\simeq1.8\times 10^{31}$~erg~s$^{-1}$) consistent with the range seen in both BW and RB spiders \citep[e.g.,][]{Koljonen2023}. The two confirmed huntsman pulsars seem to have higher $L_\mathrm{X}\gtrsim 10^{32}$~erg~s$^{-1}$, but with only two members and distance uncertainties the class properties are still unclear. The spectrum suggests that the X-ray source is relatively soft (photon index $2.6\!-\!4.9$). Most spiders are harder (index $1\!-\!2$), including the confirmed huntsman J1417 \citep[with photon index 1.6,][]{Camilo2016}. In our proposed alternative identification, X-rays could arise from an intra-binary shock between the winds of the two red giants.

    \item {\it $\gamma$-rays}. As we discussed in Section~\ref{sec:gamma-ray_variability}, J0846 was detected only briefly in 2009, but MSPs (and pulsars in general) are known to be steady $\gamma$-ray emitters. The $\gamma$-ray luminosity we estimated ($L_\gamma\simeq1.2\times 10^{34}$~erg~s$^{-1}$; Sec.~\ref{sec:gamma-ray_variability}) is consistent with the general MSP range \citep[$10^{32}\!-\!10^{34}$~erg~s$^{-1}$,][]{3PC-Smith23}. However, MSPs are also known to have curved $\gamma$-ray spectra, but \citet{Swihart2017} found no significant evidence of spectral curvature in the LAT spectrum. We also performed a curvature test on the second time bin, when the source was detected. With TS=0.0007, we confirm that there is no significant curvature in the spectrum. Again, this weakens the MSP identification, and the 2009 flare could perhaps be due to an unrelated blazar in the RoI, or a background fluctuation.
\end{itemize}

All in all, we deem the evidence for an MSP in J0846 too weak to claim it as a spider candidate, since we find no strong evidence for interactions between the two members of the binary. We propose an alternative explanation for the multi-wavelength properties of J0846: a double giant. Red giants in binaries are not common, but a few have been uncovered over the past decade \citep{Gaulme13,Rawls16,Thiemann21}. 

We note that the spectral decomposition shown in Figure~\ref{fig:J0846_SED} is not unique, and it can be sensitive to variability or systematic uncertainties on the flux densities in different bands. Alternatively, J0846 could be a non-eclipsing RS CVn system: a close binary with chromospherically active main-sequence stars \citep{Hall1976}, often detected in X-rays \citep[e.g.][]{Rengarajan1983}. Our inferred radii, however, are larger than those found in most RS CVns \citep{Toet2021}.

Our new scenario predicts a double spectroscopic binary. Future NIR spectroscopy or a reanalysis of the NIR end of the Keck-HIRES spectrum could test this and constrain more orbital parameters (e.g., $q$ will be precisely known if the radial velocity semi-amplitude of the primary, $K_1$, can be measured). The huntsman spider class is rare, with only two systems confirmed to date, and we have argued that J0846 does not belong to it.

\subsection{4FGL~J0935.3+0901: An Irradiated Redback in the Pulsar State} \label{sec:discussion_J0935}
We now turn to J0935, for which three different identifications have been proposed in the literature: a transitional MSP \citep{Wang2020}, a RB spider in the pulsar state \citep{Zheng2022} or a flaring BW candidate \citep{Halpern2022}.

Since the radio MSP has not been detected to date \citep{Zheng2022,Corcoran2023}, we do not have dynamical constraints on the companion mass. We assume a distance to J0935 of $d=1.2_{-0.5}^{+1.1}~ \mathrm{kpc}$ estimated from the Gaia-DR3 parallax measurement \citep[][but note that the parallax uncertainty is large]{Koljonen2025}. The 4FGL $\gamma$-ray spectrum is curved (4.15~$\sigma$ significance) and shows a cut-off around 1.1~GeV, consistent with an MSP identification \citep{Abdollahi20}. As is common in pulsars, the LAT $\gamma$-ray flux is quite steady (variability index 15.5, \citealt{Abdollahi20}, although \citealt{Wang2020} claimed it was variable around 2010-2013). The $\gamma$-ray luminosity is also widely consistent with that seen in MSPs \cite[$L_\gamma\simeq8\times10^{32}$~erg~s$^{-1}$ at 1.2~kpc,][]{3PC-Smith23}. The X-ray photon index ($\Gamma\simeq1.9$) and luminosity ($L_X\simeq2.5\times10^{31}$~erg~s$^{-1}$ at 1.2~kpc) are consistent with either a RB or BW in the pulsar state.

Next, we discuss the nature of the source in the context of our new optical photometry. The optical variability and changing colors presented in Section~\ref{sec:results_J0935} are consistent with an irradiated companion star in the pulsar state. We clearly detect J0935 at inferior conjunction ($\phi\simeq0$) with our LCO (1-m) observations, and find relatively low peak-to-peak amplitudes in our optical light curves. This points to a RB identification, since BWs have colder brown-dwarf like companions and often prove difficult to detect around inferior conjunction, when the cold (un-irradiated) face is visible. To estimate the spectral type and effective temperature for J0935, we corrected the observed magnitudes for interstellar reddening using the reddening value for J0935 ($E(g\!-\!r)=0.05\pm 0.02~ \mathrm{mag}$, obtained from Bayestar19 3D dust maps; \citealt{Green2019}). The $r'\!-\!i'$ color reveals a seemingly periodic irradiation pattern, suggesting our source exhibits higher temperatures around superior conjunction of the companion ($\phi\simeq0.5$; Fig.~\ref{fig:J0935_LC_Col}). From the intrinsic colors around superior conjunction ($r'\!-\!i'=0.0\pm0.1 ~\mathrm{mag}$), we estimate that $T_\mathrm{sup}$ is in the range $5750 - 7250$~K. From the intrinsic colors around inferior conjunction ($r'\!-\!i'=0.4\pm0.1 ~\mathrm{mag}$), we find $T_\mathrm{inf}$ to be in the $4000\!-\!4600$~K range (which corresponds to a K4V-K8V spectral type; \citealt{Lenz1998}; \citealt{Pecaut2013}). The latter corresponds to a flux-weighted average of all visible surface elements at inferior conjunction, presumably dominated by the base ``un-irradiated" temperature of the companion ($T_\mathrm{base}$). Thus, our $T_\mathrm{inf}$ constraints support a RB identification, since most RB companions have $T_\mathrm{base}\sim4000\!-\!6000$~K \citep{Turchetta2023}. Better color measurements along the orbit are needed to confirm these temperature estimates.

\begin{figure}[ht!]
    \centering
    \includegraphics[width=\columnwidth]{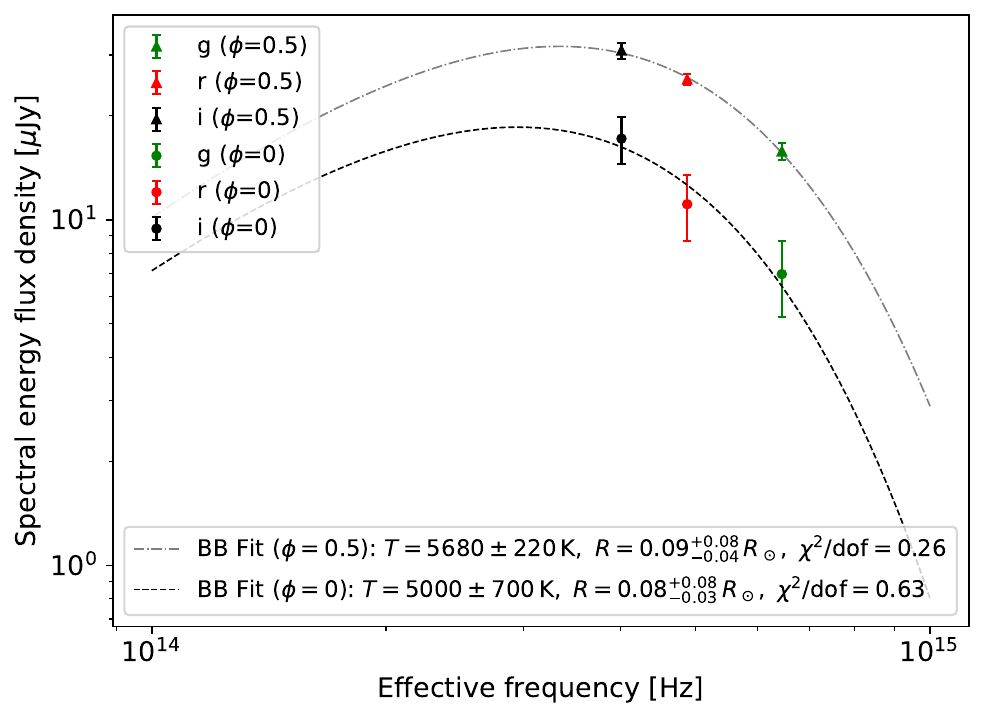}
    \caption{Spectral energy distribution of J0935 with blackbody fits at inferior and superior conjunctions of the companion. The radius and temperature are set as free parameters, while the distance of $d=1.20\ \mathrm{kpc}$ was assumed \citep{Koljonen2025}.}
    \label{fig:J0935_SED}
\end{figure}

We show the optical SED of J0935 around inferior and superior conjunction in Figure~\ref{fig:J0935_SED}, from our intrinsic (de-reddened) flux densities closest to phase $0$ and $0.5$, respectively. To fit for the inferior and superior conjunction temperatures of the companion, we use a simple blackbody model (as in Sec.~\ref{sec:sed_J0846}) and set the companion's temperature and radius as free parameters. We assume a $d=1.2_{-0.5}^{+1.1}$~kpc distance in our fits \citep{Koljonen2025}. We find $T_{BB,~0.5}=5680\pm220 ~\mathrm{K}$ at $\phi\simeq0.5$ with $\chi^2/\mathrm{dof}=0.26$, slightly lower than our $T_\mathrm{sup}$ estimate above. Around inferior conjunction ($\phi\simeq0$), we find $T_{BB,~0}=5000\pm700~ \mathrm{K}$ with $\chi^2/\mathrm{dof}=0.63$, slightly higher than the $T_\mathrm{inf}$ range derived above (we attribute these discrepancies to the difference between a stellar atmosphere and a pure blackbody spectrum). The inferred radius of the companion is $\simeq0.1_{-0.04}^{+0.08}~ \mathrm{R_{\odot}}$.

These SED fits also yield companion temperatures in the range expected for RB spiders, so we favor the RB identification of \citet{Zheng2022}. Pending a pulsar detection, we have argued that J0935 is in the pulsar (disk-free) state, so we encourage continued radio/$\gamma$-ray searches for the MSP.

As already discussed by \citet{Wang2020}, the optical emission lines could be transient and not due to the presence of an accretion disk. The emission lines observed by \citet{Wang2020} could be produced by the intra-binary shock between the pulsar wind and the material ablated from the companion star, as observed in a few confirmed spiders \citep[e.g.,][]{2019ApJ...872...42S,2021A&A...649A.120M}. We have also argued that the companion is a K4-K8 low-mass star, similar to those of known RBs. Pulsar timing (via dynamical constraints on $M_2$) or more detailed modeling of the system are needed to confirm this identification (both beyond the scope of this work).

\section{Summary and Conclusions}
We have presented the results of new optical photometry of two systems proposed as spider pulsar candidates, from robotic observations taken with the small 1-m class TJO and LCO telescopes in 2023 and 2024. We have also analyzed their multi-wavelength properties when possible (mainly the IR, X-ray and $\gamma$-ray bands).

In 2FGL~J0846.0+2820 (J0846), we presented evidence for the absence of interaction between the primary and the optical companion: the light curves indicate lack of tidal distortion and irradiation. The broadband SED shows evidence for two blackbody-like components with similar temperatures and radii, which dominate the optical ($T_2\simeq6000$~K, $R_2\simeq1.8$~R$_\odot$) and NIR ($T_1\simeq3400$~K, $R_1\simeq5.2$~R$_\odot$) bands. Based on this, and after reviewing its multi-wavelength properties, we argue that J0846 is not a bona fide pulsar candidate and propose an alternative identification: a detached double red giant. Other proposed spider candidates with little evidence for interaction may need to be revised, if the putative pulsar or its effect on the companion star remain undetected \citep{Lin24}.

In 4FGL~J0935.3+0901 (J0935), we find a light curve with one flux maximum around superior conjunction of the companion, with an accompanying increase in the optical colors and inferred temperature. We place constraints on the companion base temperature ($4000\!-\!4600$~K) which support one of the three classifications proposed for J0935: a redback spider in the pulsar state. Our results highlight the importance of finding evidence for interaction between the putative pulsar and its companion, when identifying new spider candidates.

\begin{acknowledgments}

This project has received funding from the European Research Council (ERC) under the European Union’s Horizon 2020 research and innovation program (consolidator grant agreement No. 101002352, PI: M. Linares). We thank the referee for their constructive comments and for suggesting the alternative RS CVn identification.

This publication makes use of 
observations from the Las Cumbres Observatory (LCO) global telescope network; 
data products from The Joan Oró Telescope (TJO) at the Montsec Observatory (OdM), owned by the Catalan Government and operated by IEEC;
the Catalina Sky Survey (CSS), which is funded by NASA under Grant No. NNG05GF22G;
the Two Micron All Sky Survey (2MASS), which is a joint project of the University of Massachusetts and the Infrared Processing and Analysis Center/California Institute of Technology, funded by NASA and NSF;
the Wide-field Infrared Survey Explorer (WISE), which is a joint project of the University of California, Los Angeles, and the Jet Propulsion Laboratory/California Institute of Technology, funded by NASA;.
the Pan-STARRS1 Surveys (PS1) and the PS1 public science archive, developed at the University of Hawaii's Institute for Astronomy;
{\it Fermi}-LAT data obtained from the Fermi Science Support Center (FSSC),
as well as 
data obtained from the Chandra Data Archive and the Chandra Source Catalog, both provided by the Chandra X-ray Center (CXC). The Chandra dataset is contained in the Chandra Data Collection (CDC) 596~\dataset[doi:10.25574/cdc.596]{https://doi.org/10.25574/cdc.596}.

\end{acknowledgments}

\begin{contribution}

B.D. led the development of this work, including writing and preparing the majority of the original manuscript draft, and led the optical-NIR research of J0846 (data curation, formal analysis, validation, visualization).

J.D. led the research on J0935 (data curation, formal analysis, validation, visualization) and was responsible for writing the corresponding sections in the original manuscript draft.

M.L. conceived the original idea for the project, supervised the research and reviewed, edited, and contributed to the manuscript throughout its preparation. M.L. also contributed to the data acquisition and X-ray analysis.

M.T. supervised the research and reviewed, edited, and contributed to the manuscript throughout its preparation. M.T. also supported the methodology development, formal analysis and validation.

M.S. led the high-energy research on J0846 (data curation, formal analysis, validation, visualization) and was responsible for writing the corresponding sections in the manuscript.

\end{contribution}

\section*{Data Availability} 
The dataset behind Figure \ref{fig:J0846_LC_Col}, Figure \ref{fig:J0935_LC_Col} and Figure \ref{fig:J0846_LAT_LC}, as well as the reduced spectra extracted from the Chandra observation of J0846, is publicly available at the Zenodo repository~\dataset[doi:10.5281/zenodo.22704753]{https://doi.org/10.5281/zenodo.22704753}.

\bibliography{bibtex}{}
\bibliographystyle{aasjournalv7}

\end{document}